\documentclass[longauth]{aa} % for the long lists of affiliations 
\usepackage{graphicx}
\usepackage{txfonts}
\usepackage{verbatim}
\usepackage{natbib}
\usepackage{hyperref}
\usepackage{ulem}
\newcommand{\Msun}{\mbox{$M_{\odot}$}}
\newcommand{\Rsun}{\mbox{$R_{\odot}$}}
\newcommand{\Lsun}{\mbox{$L_{\odot}$}}
\newcommand{\raiseChi}{\raise0.3ex\hbox{$\chi$}}

\begin{document}
   \title{ALMA imaging of the CO snowline of the HD 163296 disk with DCO$^+$}

\author{G. S. Mathews\inst{1} \thanks{gmathews@strw.leidenuniv.nl}
 \and  P. D. Klaassen\inst{1} 	
 \and  A. Juh\'asz\inst{1} 		
 \and  D. Harsono\inst{1,2} 	
 \and E. Chapillon\inst{3}		
 \and E. F. van Dishoeck\inst{1,4}	
 \and D. Espada\inst{5,6}			
 \and I. de Gregorio-Monsalvo\inst{5,7}	
 \and A. Hales\inst{8}		
 \and M. R. Hogerheijde\inst{1}	
 \and J.C. Mottram\inst{1}		
 \and M.G. Rawlings\inst{9}	
 \and S. Takahashi\inst{3}		
 \and L. Testi\inst{7,10}		
}

\authorrunning{G.S. Mathews et al.}

\institute{Leiden Observatory, Leiden University, PO Box 9513, 2300 RA Leiden, The Netherlands			%1	1
\and
SRON Netherlands Institute for Space Research, PO Box 9700 AV, Groningen, The Netherlands			%2	2
\and
Academia Sinica Institute of Astronomy and Astrophysics (ASIAA), P.O. Box 23-141, Taipei 10617, Taiwan	%4	3
\and 
Max-Planck-Institut f\"ur Extraterrestrische Physik, Giessenbachstrasse 1, 85748 Garching, Germany		%6	4
\and
National Astronomical Observatory of Japan (NAOJ), 2-21-1 Osawa, Mitaka, Tokyo 181-8588, Japan		%3	5
\and
NAOJ Chile Observatory																	%7	6
\and
European Southern Observatory, Karl Schwarzschild Str 2, D-85748 Garching bei M\"unchen, Germany		%8	7
\and 
Joint ALMA Observatory (JAO), Alonso de Cordova 3107, Vitacura, Santiago, Chile						%5	8
\and
National Radio Astronomical Observatory (NRAO), 520 Edgemont Road, Charlottesville, VA 22903, USA	%9	9
\and
INAF - Osservatorio Astrofisico di Arcetri, Largo E. Fermi 5, 50125, Firenze, Italy						%10	10
}
 
   \date{Accepted to A\&A, July 11, 2013}

% \abstract{}{}{}{}{} 
% 5 {} token are mandatory
 
  \abstract
   {The high spatial resolution and line sensitivity of the Atacama Large Millimeter/submillimeter Array (ALMA) opens the possibility of resolving emission from molecules in large samples of circumstellar disks.  With an understanding of the conditions under which these molecules can have high abundance, they can be used as direct tracers of distinct physical regions.  In particular, DCO$^+$ is expected to have an enhanced abundance within a few Kelvin of the CO freezeout temperature of 19 K, making it a useful probe of the cold disk midplane.}
   {We aim to use line emission from DCO$^+$ to directly resolve the CO `snowline' -- the region at which the gas-phase CO abundance drops due to freezeout -- and determine the temperature boundaries of the region of DCO$^+$ emission in the \object{HD 163296} disk.  This will serve as a test of deuteration models based on enhanced formation of the parent molecule H$_2$D$^+$ and a direct probe of midplane disk structure and ionization.}
   {We compare ALMA line observations of HD 163296 to a grid of models based on the best fit physical model of \cite{Qi:2011}.  We vary the upper- and lower-limit temperatures of the region in which DCO$^+$ is present as well as the abundance of DCO$^+$ in order to fit channel maps of the DCO$^+$ $J$=5$-$4 line.  To determine the abundance enhancement compared to the general interstellar medium, we carry out similar fitting to HCO$^+$ $J$=4$-$3 and H$^{13}$CO$^+$ $J$=4$-$3 observations.}
   {ALMA images show centrally peaked extended emission from HCO$^+$ and H$^{13}$CO$^+$.  DCO$^+$ emission lies in a resolved ring from $\sim$110 to 160 AU.  The outer radius approximately corresponds to the size of the CO snowline as measured by previous lower resolution observations of CO lines in this disk.  The ALMA DCO$^+$ data now resolve and image the CO snowline directly. }  
   {In the best fitting models, HCO$^+$ exists in a region extending from the 19 K isotherm to the photodissociation layer with an abundance of $3\times10^{-10}$ relative to H$_2$.  DCO$^+$ exists within the 19--21 K region of the disk with an abundance ratio [DCO$^+$] / [HCO$^+$] = 0.3.  This represents a factor of 10$^4$ enhancement of the DCO$^+$ abundance within this narrow region of the \mbox{HD 163296} disk.  Such a high enhancement has only previously been seen in prestellar cores.  The inferred abundances provide a lower limit to the ionization fraction in the midplane of the cold outer disk ($\gtrsim4\times10^{-10}$), and suggest the utility of DCO$^+$ as a tracer of its parent molecule H$_2$D$^+$.}

   \keywords{
                Stars: pre-main sequence; (Stars:) planetary systems: protoplanetary discs; submillimeter: stars, Stars: individual: HD 163296
               }

   \maketitle
%
%________________________________________________________________

\section{Introduction}

Circumstellar disks of gas and dust are the environments in which planets form, as well as many molecules which may be transported to planets.  
Understanding the density and temperature structure of these disks, in particular the dense midplane, is necessary to understand the dynamic and chemical processes involved \citep{Williams:2011}.  
However, directly probing the midplane is difficult.  Many of the most abundant molecules, such as CO and HCO$^+$, are present throughout much of the disk but have high opacity.  Optically thin isotopologues such as C$^{18}$O or H$^{13}$CO$^+$ may have detectable emission from the midplane, but it is confused by emission from higher layers.  In addition, much of the midplane becomes unobservable as CO and other tracers freeze onto grains due to low temperatures at large radii.

DCO$^+$ has been suggested as an excellent tracer of the disk midplane near the freezeout region of CO because the decline in temperature provides an ideal environment for gas-phase DCO$^+$ formation.  It is formed in the gas phase primarily by the transfer of a deuterium atom from the H$_2$D$^+$ ion to a CO molecule \citep{Wootten:1987}.  

Chemical models predict the enhanced formation of the parent molecule H$_2$D$^+$ at low temperatures from the reaction 
\begin{equation}
\rm{H}_3^+ + HD \Leftrightarrow H_2D^+ + H_2 + \Delta E, 
\end{equation}
where $\Delta$E $\sim$220 K% due to the energy barrier to the back reaction H$_2$D$^+$ + H$_2$ 
~\citep[][]{Roberts:2000}.  This effect is further enhanced by the reduction in the H$_2$ ortho/para ratio at low temperatures, where the energy difference between the ortho ground state ($J=1$) and para ground state ($J=0$) otherwise provides internal energy for the back reaction \citep[$\Delta\rm{E}\sim170 ~\rm{K}$,][]{Maret:2007fj,Pagani:2009}.  

The parent molecule of H$_2$D$^+$, H$_3^+$, %, a parent molecule for H$_2$D$^+$, 
~is effectively destroyed by CO \citep{Jorgensen:2004}, which can remain in the gas phase down to a temperature of $\sim20$ K.  Thus a reduction in the CO abundance will make HD the preferred destroyer of H$_3^+$ and lead to an enhancement of H$_2$D$^+$.  The increase in H$_2$D$^+$ formation, in turn, is expected to enhance the formation of DCO$^+$.  

However, CO is also one of the parent molecules of DCO$^+$, so while DCO$^+$ formation is aided by a decrease in CO abundance, it will also be constrained at temperatures where too little CO remains in the gas phase.  A balance between low temperatures and CO abundance is required to enhance DCO$^+$ abundance.

\begin{figure*}[th]
   \centering
%\vspace{-1cm}  \hspace{-0.5cm}    \includegraphics[width=1.05\textwidth]{figures/mom0_all_lines/mom0_all_lines.eps}
\vspace{-1cm}  \hspace{-0.5cm}    \includegraphics[width=1.05\textwidth]{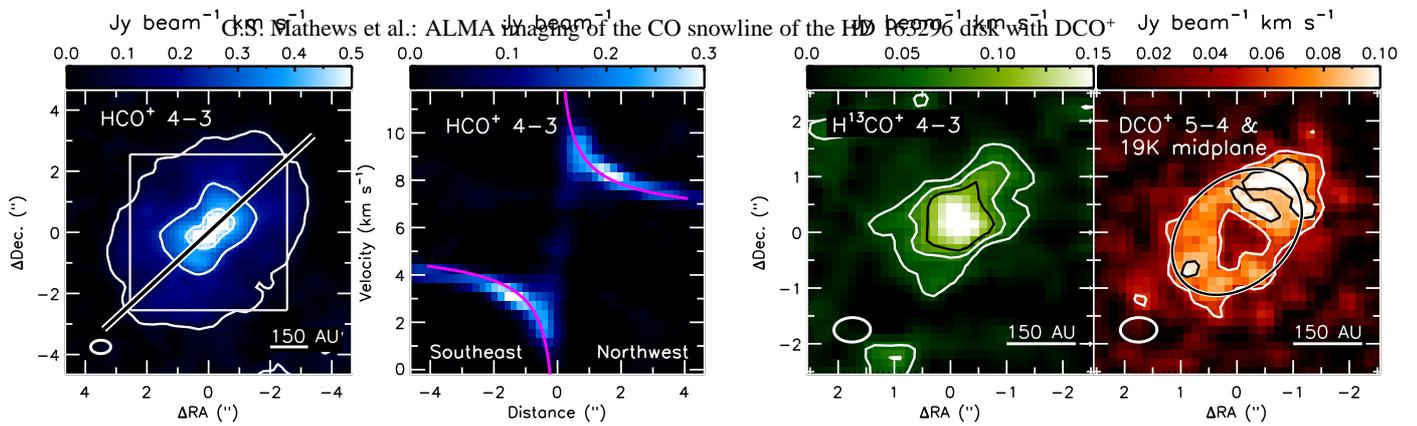}
\vspace{-0.65cm}      \caption{\textbf{Left:} Integrated emission map of HCO$^+$ (blue) from 0--12 km s$^{-1}$ with contours showing 3, 10, and 17$\sigma$ emission ($\sigma$ = 25 mJy beam$^{-1}$ km s$^{-1}$).  The white-bordered black line shows the positions from which we constructed the position-velocity diagram (middle-left), and the gray box shows the smaller region for which we show H$^{13}$CO$^+$ and DCO$^+$ line emission (middle-right and right).  \textbf{Middle-left:} The HCO$^+$ position-velocity diagram.  The magenta curves show the Keplerian velocity profile for an assumed stellar mass of 2.3 \Msun, disk inclination of 44\degr, and systemic velocity of 5.8 km s$^{-1}$.  \textbf{Middle-right:}  Integrated emission map of H$^{13}$CO$^+$ (green) from 0.8 --10.4 km s$^{-1}$ with contours showing 3, 5, and 7$\sigma$ emission ($\sigma$ = 14 mJy beam$^{-1}$ km s$^{-1}$).  \textbf{Right:} Integrated emission map of  DCO$^+$ (red) from 0.8 --10.4 km s$^{-1}$ with contours showing 3 and 5 $\sigma$ emission ($\sigma$ = 18 mJy beam$^{-1}$ km s$^{-1}$).  The 19 K contour of the midplane temperature at a radius of 155 AU in the model disk is overlaid on the DCO$^+$ emission using a white-bordered black ellipse.  In the maps of integrated emission, the $\sim$0\farcs65 $\times$ 0\farcs44 beam is shown in the lower left, while a scale is shown in the lower right. }
       \label{fig:allLines}
\end{figure*}

These factors have led to models predicting that the DCO$^+$ abundance will be enhanced in a narrow range of temperatures of a few Kelvin near the CO freezeout temperature of $\sim20$ K, which successfully explain emission from pre-stellar cores \citep[e.g.,][]{Caselli:1999, Pagani:2012qy}.  In disks, line emission from DCO$^+$ is then expected to trace the rise in formation of its parent molecule, H$_2$D$^+$ at increasing radii.  H$_2$D$^+$ serves as the basis of many deuteration processes, and is one of the principal ions in cold dense gas.  DCO$^+$ emission will not rise indefinitely, instead it will form a ring which is truncated at the midplane $\sim$20 K isotherm, the CO `snowline' where the gas phase CO abundance drops due to freezeout onto grains \citep[e.g.,][]{Aikawa:2002, Willacy:2007}.  Tracing the CO snowline is of importance to theories of planet formation because of the increase in the surface density of solids which can accelerate the growth of larger bodies.   

While ions such as HCO$^+$ and DCO$^+$ are expected to trace low temperature conditions within the disk, they can also provide limits on the ionization fraction \citep[e.g.,][]{Oberg:2011c}, a key parameter governing accretion in the disk.  At ionization fractions ($x_{\rm{ion}}$) less than $10^{-13}$, the magneto-rotational instability that drives viscous accretion breaks down due to a lack of conductivity.  This creates `dead zones' \citep{Gammie:1996zr}, which in turn provide one mechanism of enhancing planet formation by the generation of high pressure vortices that trap large dust grains, thus increasing grain growth efficiency \citep[e.g.,][]{Barge:1995ve,Varniere:2006gf}.  Direct assessment of the ionization fraction will serve to test recently proposed mechanisms by which cosmic rays, one of the primary ionization mechanisms, might be excluded from a circumstellar disk by the young star's stellar winds and magnetic field \citep{Cleeves:2013lr}.

Single dish observations have previously revealed emission from DCO$^+$ towards two T Tauri stars, TW Hya  \citep{2003A&A...400L...1V, Guilloteau:2006}, with interferometric observations suggesting a ringlike morphology for DCO$^+$ in TW Hya \citep{Qi:2008,Oberg:2012a}.  Comparison with HCO$^+$ has indicated a disk averaged DCO$^+$ / HCO$^+$ abundance ratio of $\sim$0.04 \citep[TW Hya,][]{2003A&A...400L...1V}, which is a factor of 1000 higher than the interstellar medium [D]/[H] abundance of $\approx10^{-5}$ \citep{Watson:1976}.  Even higher abundances have been seen in dark clouds such as L1544 \citep[DCO$^+$/HCO$^+$ = 0.12,][]{Caselli:1999}.  Later observations with the Sub-Millimeter Array (SMA) have detected DCO$^+$ in several additional disks, though not with sufficient resolution to see the details of the morphology \citep{Oberg:2010a,Oberg:2011b,Oberg:2011c}.

Nearby gas-rich disks provide ideal sites for examination of the cold disk midplane using DCO$^+$.  One such system is the Herbig Ae star \mbox{HD 163296} \citep[$\alpha$ = 17h56m21.29s, $\delta$ =  $-$21\degr57'21\farcs87, J2000,  spectral type A1,][]{Houk:1988}, which is relatively nearby \citep[122 pc,][]{1997A&A...323L..49P} and young \citep[4 Myr,][]{van-den-Ancker:1998}.  It has long served as a prototype for studies of gas and dust rich protoplanetary disks \citep[e.g.,][]{Mannings:1997}.  Previous studies have determined that it has a relatively massive disk (0.07--0.08 \Msun) and a gas-to-dust ratio from 100 to 150 \citep[e.g.,][]{Qi:2011,Tilling:2012}.  The disk is observed at an inclination of 44\degr, and position angle of $\sim$133\degr ~East-of-North.  Due to its high luminosity ($\sim40$ \Lsun ), the 20 K region is at a large radius which can be resolved with science verification data from the Atacama Large Millimeter/submillimeter Array (ALMA).  The massive disk has a high total line-of-sight column density of DCO$^+$ and the intermediate inclination means that vertical and radial spatial information can be obtained, as well as kinematic information.  

\citet[henceforth Q11]{Qi:2011} measured the radius of the CO snowline in the disk around \mbox{HD 163296} by fitting high spectral resolution visibility data from the SMA with models of $^{13}$CO emission incorporating a sharp drop in abundance as a function of distance from the star.   They estimated that the CO snowline of \mbox{HD 163296} lies at a distance of 155 AU, corresponding to a midplane temperature of 19 K in their physical model.

 In this paper, we present resolved imaging of DCO$^+$ line emission from \mbox{HD 163296} and examine the utility of DCO$^+$ as a tracer of disk structure. The paper is structured as follows.  In \S \ref{sec:obs}, we describe the observations and data reduction, and present our results in \S \ref{sec:results}.  We compare the disk properties to models in \S \ref{sec:modeling} and explore the implications for further disk studies in \S \ref{sec:discussion}.  We summarize our findings in \S \ref{sec:conclusions}.

\section{Observations}
\label{sec:obs}

Band 7 observations of \mbox{HD 163296} were carried out on June 9, 11, 22, and July 6, 2012 for a total on-source time of 59.4 minutes as part of the ALMA science verification program 2011.0.000010.SV.
~ Median precipitable water vapor levels (pwv) were 1.8, 0.7, 0.3, and 0.7 mm, respectively.  
The ALMA array included between 17 and 20 12m diameter antennas and baselines from 13m to 402m.  The four correlator bands were centered on the CO $J$=3-2, HCO$^+$ $J$=4$-$3, H$^{13}$CO$^+$ $J$=4$-$3, and DCO$^+$ $J$=5$-$4 transitions at 345.796, 356.734, 346.998, and 360.170 GHz, with channel widths of 122.1, 122.1, 244.1, and 30.5 kHz, and total bandwidths of 468.8, 468.8, 937.5, and 117.2 MHz, respectively.  
For gain, bandpass, and flux calibration, we used J1733-130, J1924-292, and Neptune, respectively.

\begin{table}[]
     \center
      \caption[]{Observational and modeling results.  Each species and transition is given along with the rms in emission-free channels with a velocity width of 0.4 km s$^{-1}$.  The third and fourth columns show the line flux integrated over the disk, and the flux of the adopted model.}
      \label{tab:results}
%       \resizebox{18.5cm}{!} { 

         \begin{tabular}{l lll  }
            \hline\hline
            \noalign{\smallskip}
            Line   					&  RMS						&  Flux			&	Model flux		\\
                           					&    (mJy beam$^{-1}$) 	 		&  (Jy km s$^{-1}$) 	& 	(Jy km s$^{-1}$)\\
            \noalign{\smallskip}
            \hline
            \noalign{\smallskip}	
		HCO$^+$ ~~~$J$=4$-$3		&  11.3						&  18.7$\pm$0.7	&  26.0	\\ 	% chan 30 - 60      	spw1
		H$^{13}$CO$^+$ $J$=4$-$3	&   ~~8.2					&  ~~1.0$\pm$0.2	&  ~~1.4	\\	% chan 32 - 56  						spw3		
		DCO$^+$ ~~~$J$=5$-$4		&   13.1						&  ~~2.2$\pm$0.4	&  ~~1.7	\\	% chan 33 - 56							spw0
           \noalign{\smallskip}
            \hline
%           \multicolumn{7}{l}{\tablefoottext{a}{}}
         \end{tabular}
%       }	% end \resizebox
           \\
%           \tablefoottext{a} Calibration uncertainty 40\%		%  \tablefootmark{a} 	
%           \tablebib{(1) Dunkin et al., \cite{dunkin};  }
        
 \end{table}

%  Maximum observable size ~ lambda / L, where L is the shortest baseline
%  baseline lengths
%  [('CM04-CM05', 13.080350393257904),
%  ('DA43-DV17', 402.27026849954729)]

%  au.getBaselineLengths('HD163296.spw0.ms', sort=True)
%  au.plotPWV('uid___A002_X4267ef_X358.ms', figfile='', plotrange=[0,0,0,0], clip=True)
%  au.timeOnSource('uid___A002_X4267ef_X358.ms')

We used the Common Astronomy Software Applications package \citep[CASA,][]{McMullin:2007} to carry out calibration and imaging of the visibility datasets.  Self-calibration was carried out for both phases and amplitudes using line free channels.  We used the CLEAN algorithm with natural weighting and circular clean masks with diameters of $\sim$9'', 5'', and 5'' to produce channel maps of the HCO$^+$, H$^{13}$CO$^+$, and DCO$^+$ lines, respectively.   The image cubes have a velocity resolution of 0.4 km s$^{-1}$ and pixel size of 0\farcs15 in each data cube.  The synthesized beam size is $\sim$0\farcs65 $\times$ 0\farcs44, at a position angle of $\sim$90\degr.  

In this paper we examine the emission from HCO$^+$ and its isotopologues, with a focus on DCO$^+$.  Discussion of the CO line emission from the disk may be found in de Gregorio-Monsalvo et al., (in press) and in Chapillon et al., (in prep), while \cite{Klaassen:2013qy}, discuss CO line emission associated with the disk wind.

\section{Results}
\label{sec:results}

 In Table \ref{tab:results}, we list the rms noise limits found in map regions and channels showing no emission, as well as the spatially and spectrally integrated emission in each line.  We also present the line fluxes of our adopted model (discussed below).  Line emission is detected at velocities between 0 and 12 km s$^{-1}$ in HCO$^+$, and at velocities of 0.8 to 10.4 km s$^{-1}$ in H$^{13}$CO$^+$ and DCO$^+$.  All three lines have a central LSR velocity of 5.8 km s$^{-1}$, consistent with previous observations \citep[e.g.,][Q11]{Mannings:1997}.  In Figure \ref{fig:allLines}, we show the integrated maps of the emission from all three isotopologues in the velocity ranges given above.

\subsection{Morphology}
The HCO$^+$ emission exhibits a smooth spatial profile that peaks at the stellar position. We fit an ellipse to the 3$\sigma$ contour of the integrated emission map, and find that it has major and minor axes of 7\farcs8 and 5\farcs3 and a position angle of 133\degr ~east of north.  Under the assumption of a flat disk these correspond to a disk with an outer radius of 475 AU observed at an inclination of 47\degr, comparable to previous inclination determinations.
The position-velocity diagram in Figure \ref{fig:allLines} shows that the line-of-sight velocities along the disk major axis exhibit a Keplerian velocity profile.

~H$^{13}$CO$^+$ emission is only detected near the center of the disk, with the 2$\sigma$ contour extending 1\farcs4 from the disk center (170 AU).

DCO$^+$ emission is not detected at the stellar position but is confined to an annulus in the integrated emission map with outer major and minor axes of 3\farcs3 and 2\farcs2, and inner major and minor axes of 1\farcs4 and 1\farcs0.  This suggests DCO$^+$ emission originates in a ring with a central radius of $\sim140$ AU and a width less than 110 AU (i.e. extending from a radius of at least 95 AU to less than 195 AU).  
~This approximately corresponds to the CO snowline found by Q11 from lower resolution SMA observations of CO and its isotopologues, a relation we explore in more detail below.  We show the midplane 19 K contour at a radius of 155 AU from our adopted model (see \S \ref{sec:modeling}) in Figure \ref{fig:allLines}.  There is an apparent peak in emission to the northwest.  However, in this work we address the large scale structure, and do not attempt to model variations within that structure.

\subsection{Estimated column densities}
\label{sec:column_density}
H$^{13}$CO$^+$ is expected to have an abundance a factor of 75 less than HCO$^+$, corresponding to the $^{12}$C/$^{13}$C abundance ratio in the interstellar medium \citep{Frerking:1982}.   We assume that the H$^{13}$CO$^+$ $J$=4$-$3 line is optically thin and can be used to estimate the median molecular column density in the disk.  The H$^{13}$CO$^+$ $J$=4$-$3 / HCO$^+$ $J$=4$-$3 intensity ratio for positions and velocities where H$^{13}$CO$^+$ is detected has a median value of 0.18, confirming the high optical depth of HCO$^+$ with $\tau\approx16$ under the assumption of LTE.

We use the online RADEX 1D non-LTE radiative transfer tool\footnote{http://www.sron.rug.nl/\textasciitilde{}vdtak/radex/radex.php} \citep{van-der-Tak:2007} to make an initial estimate of the column density.  
Assuming a kinetic temperature $T_{\rm{K}}$ = 40 K and gas density $n_{\rm{H}_2}$ = 10$^{7}$ cm$^{-3}$ (intermediate values from the disk model of Q11), and assuming $\Delta\rm{V}$ = 1 km s$^{-1}$, we adjust the HCO$^+$ and H$^{13}$CO$^+$ column densities until the model intensity ratios match the median observed intensity ratio.  We find line of sight column densities of 2.1$\times10^{14}$ cm$^{-2}$ and 2.8$\times10^{12}$ cm$^{-2}$ for HCO$^+$ and H$^{13}$CO$^+$, respectively.  Correcting for the disk inclination, these are vertical column densities of 1.5$\times10^{14}$ cm$^{-2}$ and 2.0$\times10^{12}$ cm$^{-2}$.

The column density sensitivity of our observations is estimated using RADEX, as well.  For each transition, we convert the 3$\sigma$ limits in our maps of integrated line emission to antenna temperature.  We then assume gas densities and temperatures typical of the expected emission regions.  For HCO$^+$ and H$^{13}$CO$^+$, we assume the same $T_{\rm{K}}$ and $n_{\rm{H}_2}$ values as above, and for DCO$^+$ we assume $n_{\rm{H}_2}$ = 10$^9$ cm$^{-3}$ and $T_{\rm{K}}$ = 19 K (values from the midplane CO snowline of Q11).  Our integrated maps have approximate column density sensitivities of $1.7\times10^{12}$, $0.7\times10^{12}$, and $2.5\times10^{12}$ cm$^{-2}$ for HCO$^+$, H$^{13}$CO$^+$, and DCO$^{+}$, respectively.  These estimates do not reflect the variation in density and temperature across the regions in which these molecules exist, factors that require more detailed radiative transfer modeling as described in the following section.

\section{Modeling}
\label{sec:modeling}

\subsection{Physical structure}

Analysis of the DCO$^+$ spatial distribution requires a physical model which specifies the temperature and gas density as a function of both radius and height in the disk.  We use a parametric disk model to approximate the best-fit model of Q11, with parameters adjusted to fit the spectral energy distribution and line fluxes for CO and its isotopologues in the literature.  The details of the density and temperature structure of our disk model, as well as our radiative transfer models and ray-tracing, can be found in Appendix \ref{sec:model_details}.  
Simulated observations of our models, mimicking the $(u,v)$ coverage of the data, were carried out using the built-in simulation tasks of CASA.  

We used this SED-constrained model as the basis for constructing a grid of HCO$^+$, H$^{13}$CO$^+$, and DCO$^+$ models, with energy levels, transition frequencies, and Einstein A coefficients taken from the Leiden Atomic and Molecular Database\footnote{http://home.strw.leidenuniv.nl/\textasciitilde{}moldata/} \citep{Schoier:2005}.  We manually varied the isotopologue abundances, and in the case of DCO$^+$ the region in which it is abundant (discussed below), and compared the resulting channel maps to the observations.

\subsection{Molecular abundances}

The HCO$^+$ and H$^{13}$CO$^+$ abundances (\raiseChi$_{\rm{HCO^+}}$ and \raiseChi$_{\rm{H^{13}CO^+}}$, respectively) are taken to be constant within the region populated by the parent CO molecule.  This region is bounded on the upper side by photodissociation \citep[i.e., below the height where a hydrogen column density of $N_{\rm{H_2}} = 2\times10^{21}$ cm$^{-2}$ is reached, adopted from the translucent-cloud modeling of][]{Visser:2009a} and on the lower side by the 19 K CO freezeout zone (Q11).  Outside this region we treat the abundance as zero.  We explored HCO$^+$ abundances relative to H$_2$ ranging from 10$^{-11}$ to 10$^{-5}$ in logarithmic steps of 0.5 (corresponding to factors $\sim$3) and setting the H$^{13}$CO$^+$ abundance to 1 / 75 that of HCO$^+$.  We find a best simultaneous match to the HCO$^+$ and H$^{13}$CO$^+$ integrated intensities and channel maps with a HCO$^+$ abundance of $3\times10^{-10}$, and a corresponding H$^{13}$CO$^+$ abundance of $4\times10^{-12}$.  These constant abundances are adopted throughout this work.

We carried out a similar manual fit to the DCO$^+$ $J$=5$-$4 emission.  Our DCO$^+$ models include the same assumption of a uniform abundance (\raiseChi$_{\rm{DCO^+}}$) within the region of DCO$^+$ enhancement, and an abundance of zero outside of this region.  However, the region in which DCO$^+$ is enhanced is defined by both high temperature and low temperature cutoffs, $T_{\rm{high, DCO^+}}$ and $T_{\rm{low, DCO^+}}$.  
~Above $T_{\rm{high, DCO^+}}$, DCO$^+$ cannot form in large quantities due to the destruction of H$_3^+$ and H$_2$D$^+$ by gas phase CO and ortho-H$_2$, respectively.  Below $T_{\rm{low, DCO^+}}$ it cannot form due to a lack of gas phase CO.  We find that a \raiseChi$_{\rm{DCO^+}}$ of 1$\times$10$^{-10}$ ~at temperatures between $T_{\rm{high, DCO^+}}$ = 21 K and $T_{\rm{low, DCO^+}}$ = 19 K provides the best fit to both the total flux and to channel maps.  We compare fluxes from our adopted model to the observed fluxes in Table \ref{tab:results}.

\begin{figure}
   \centering
   \includegraphics[angle=0, width=\columnwidth]{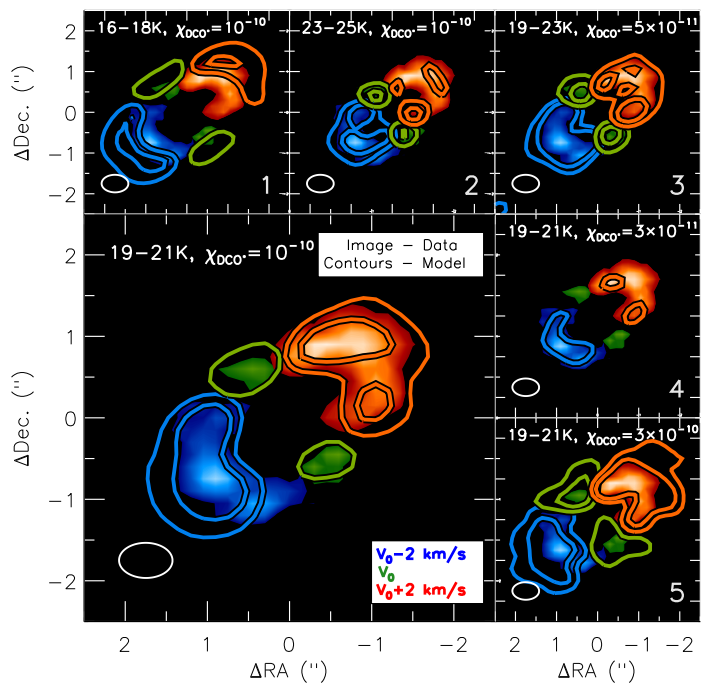}
      \caption{Comparison of integrated intensities of our DCO$^+$ models (contours, 3 and 5 $\sigma$, $\sigma$=15 mJy beam$^{-1}$ km s$^{-1}$) to the observations (image, colors show intensities from 3$\sigma$).  Colors indicate the velocities included in constructing the zeroth moment maps.  Clockwise from the upper left, small figures show: 1) decreasing $T_{\rm{low, DCO^+}}$ and $T_{\rm{high, DCO^+}}$ by 3 K, 2) raising $T_{\rm{low, DCO^+}}$ and $T_{\rm{high, DCO^+}}$ by 4 K, 3) broadening the temperature range from 2 K to 4 K, with \raiseChi$_{\rm{DCO^+}}$ halved to maintain the same total DCO$^+$ $J$=5$-$4 line flux, 4) reduction of \raiseChi$_{\rm{DCO^+}}$ by a factor of 3, and 5) raising \raiseChi$_{\rm{DCO^+}}$ by a factor of 3.}
         \label{fig:DCOp_chanmap}
   \end{figure}

Figure \ref{fig:DCOp_chanmap} presents the comparison between our best fit DCO$^+$ model and the observations in the large lower left panel, for 2 km s$^{-1}$ wide channel maps centered at 4, 6, and 8 km s$^{-1}$, respectively.  We simultaneously display these maps in blue, green, and red scaled uniformly above 3 $\sigma$, and overlay the 3 and 5 $\sigma$ model contours in the same velocity ranges (in these broad channel maps, $\sigma=$15 mJy beam$^{-1}$ km s$^{-1}$).  

The five small panels of Figure \ref{fig:DCOp_chanmap} show how the emission varies if we change the temperature range and abundance of DCO$^+$.  The panels show:  {1)} if we decrease the temperature bounds by 3 K or {2)} increase them by 4 K, {3)} if we broaden the temperature range by a factor of two (while halving the abundance to maintain approximately the same integrated emission), or if we {4)} decrease the abundance by a factor of 3, or {5)} increase it by a factor of 3.  These effects highlight the  relation between disk properties and observables: the temperature range in which DCO$^+$ has high abundance primarily sets the observed spatial region of emission, while the abundance adjusts the strength of that emission.  

\begin{comment}
In the surrounding panels, we show how the DCO$^+$ emission varies with changes in $T_{\rm{low, DCO^+}}$, $T_{\rm{high, DCO^+}}$, and \raiseChi$_{\rm{DCO^+}}$.  Clockwise from the upper left, these illustrate: 
\begin{enumerate}
\item decreasing $T_{\rm{low, DCO^+}}$ and $T_{\rm{high, DCO^+}}$ by 3 K, 
\item raising the $T_{\rm{low, DCO^+}}$ and $T_{\rm{high, DCO^+}}$ by 4 K, 
\item broadening the temperature range from 2 K to 4 K, with \raiseChi$_{\rm{DCO^+}}$ halved to roughly maintain the same total DCO$^+$ $J$=5$-$4 line flux, 
\item reduction of \raiseChi$_{\rm{DCO^+}}$ by a factor of 3, and 
\item raising \raiseChi$_{\rm{DCO^+}}$ by a factor of 3.  
\end{enumerate}
These indicate the parameter bounds at which the models are worse than our adopted model.  
\end{comment}

\begin{figure}
   \centering

   \includegraphics[angle=0, width=\columnwidth]{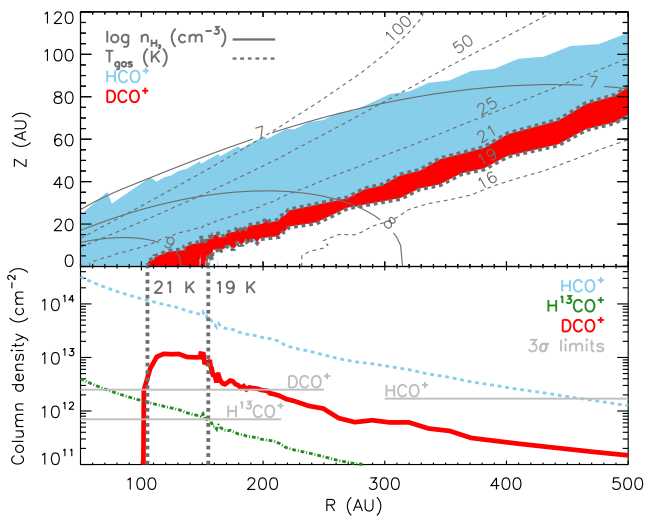}
      \caption{\textbf{Top:} Regions of non-zero HCO$^+$ (blue) and DCO$^+$ (red) abundance, overlaid with contours indicating the hydrogen density (solid gray lines) and temperature (dashed gray lines).  The 19 and 21 K contours are shown with a thick dark gray line.  The region containing HCO$^+$ is bounded on the upper and lower sides by CO photodissociation and by freezeout at 19 K, respectively.  The DCO$^+$ region is bounded by the temperature range 19 -- 21 K. ~\textbf{Bottom:} The column density of HCO$^+$ and its isotopologues (colored lines), with the 19 K CO snowline indicated by the thick dark gray dashed line.  The broad plateau of the DCO$^+$ column density from $\sim$110--150 AU explains the ring morphology seen in line emission.  The 3$\sigma$ line sensitivities (from \S \ref{sec:column_density}) are shown with the horizontal gray lines.}
         \label{fig:abundance}
   \end{figure}

In Figure \ref{fig:abundance}, we illustrate the HCO$^+$ and DCO$^+$ regions of our adopted model and the molecular column density as a function of radius in the disk.  Due to the increasing physical thickness of the DCO$^+$ region, decreasing gas density, and our assumption of a constant abundance, the DCO$^+$ column density is relatively constant from 110 to 150 AU, corresponding to the brightest region of observed emission.  

\section{Discussion}
\label{sec:discussion}

\subsection{Why a ring?}

The data constrain DCO$^+$ emission to a narrow ring near the previously inferred CO snowline, and our models constrain the physical parameters of this region.  Three factors combine to restrict the detection of DCO$^+$ to a narrow ring:  

\begin{enumerate}
\item As the temperature decreases to 21 K, the CO abundance begins to drop due to freezeout and allows the effective formation of one of the  parent molecules of DCO$^+$, H$_2$D$^+$.  In addition, the ortho/para ratio of H$_2$ drops to $\sim10^{-3}$, inhibiting the back reaction of H$_2$D$^+$ with H$_2$.  These low temperatures are first reached at a radius of $\sim110$ AU, and the DCO$^+$ column density will rise as its abundance becomes enhanced (Figure \ref{fig:abundance}, 21 K contour).    

\item Where the temperature drops below 19 K (i.e. at radii greater than 155 AU in the midplane), the other parent molecule of DCO$^+$, CO, will be largely removed from the gas phase due to freezeout.  This will in turn reduce the DCO$^+$ column density (Figure \ref{fig:abundance}, 19 K contour).  The temperature bounded region is present at larger radii above the midplane, but at these heights, DCO$^+$ emission faces a third constraint. 

\item The gas density, $n_{\rm{H}_2}$, decreases with height above the midplane, in turn reducing the column density of DCO$^+$.  This is illustrated by comparing the top and bottom panels of Figure \ref{fig:abundance}.  The column density decreases at the CO snowline at 155 AU (bottom, vertical dashed gray line).  This is the radius where the 19 K temperature contour rises above the midplane (top, thick dashed gray line).  As the 19 K isotherm rises in the disk, the gas density, and the corresponding column densities, decrease (top, solid contours).  

\end{enumerate}

In all of these aspects, the DCO$^+$ chemistry in the {HD 163296} disk is consistent with the chemistry used to explain the emission from deuterated molecules in dark clouds and prestellar cores.

At low optical depths and nearly constant temperature, the integrated intensity will be an approximately linear function of column density.  We can infer from Figure \ref{fig:abundance} that the outer edge of the broad peak of DCO$^+$ emission traces the CO snowline in the midplane, while emission at larger radii traces a layer of constant temperature that rises from the midplane of the disk.  Under our enrichment scenario, the inner edge indicates where the formation rate of H$_2$D$^+$ rises.

In our model, the column density of DCO$^+$ drops by a factor of $\sim$4 from 150 to 200 AU.  In the current observations, the peak intensity is 6$\sigma$, suggesting that only a factor of 3 improvement in sensitivity is needed to probe this disk structure to a radius of $\sim$200 AU.  With twice as many antennas, the full ALMA array will be able to accomplish such an observation with only $\sim$2 hours integration time.

\subsection{Probing the vertical structure}

Models with a factor of 3 increase in column density - which could be achieved either by increasing the relative abundance or the overall gas density - lead to much larger structures being detectable in the model images (see panels 4 and 5 of Figure \ref{fig:DCOp_chanmap}).  This is due to an increase in the column density at larger radii in the disk.  As this is also the region where the 19 K isotherm rises from the midplane, this suggests the utility of DCO$^+$  not just as a probe of the regions of enhanced H$_2$D$^+$ formation and the midplane CO snowline, but also with higher sensitivity observations a probe of the vertical disk structure.  

In combination with detailed chemical models, this provides constraints on the degree of vertical mixing, as DCO$^+$ that moves upward from this region will be destroyed, and that which moves down in the disk will freeze out on grains.

\subsection{Sensitivity to T$_{high}$ and T$_{low}$}

At a radius of 155 AU, the disk midplane temperature changes at a rate of $\approx$1 K per 20 AU.  Small changes in the modeled DCO$^+$ temperature range then correspond to large changes in the physical emission region.  The resolution of the beam, 0\farcs44, corresponds to about 50 AU at the distance of \mbox{HD 163296}, therefore differences of 2.5 K in $T_{\rm{high, DCO^+}}$ and $T_{\rm{low, DCO^+}}$ will lead to resolvable differences in the radial distribution of line emission.  The center of the temperature range can decrease by less than 3 K and increase by less than 4 K (see panels 1 and 2 of Figure \ref{fig:DCOp_chanmap}).  The width of the temperature range in which DCO$^+$ exists cannot be much larger than 2 K, as the ring of emission becomes too wide (see panel 3 of Figure \ref{fig:DCOp_chanmap}).  Changes in the position of emission in the disk will also lead to changes in the line of sight velocity, with the velocity changing on a scale of $\approx$0.2 km s$^{-1}$ per 20 AU.  $T_{\rm{high, DCO^+}}$ and $T_{\rm{low, DCO^+}}$ cannot shift far from $\sim$20 K %(corresponding to a midplane distance of $\sim$150 AU) 
as the ring both changes in size and the emission shifts to different velocities.

\subsection{Enhancement of DCO$^+$}

While DCO$^+$ is restricted to a much smaller vertical area of the disk than HCO$^+$, the regions in which these ions are abundant overlap near the CO snowline.  Under our assumption of a uniform HCO$^+$ abundance within the region from the CO snowline to the CO photodissociation limit at $N_{\rm{H_2}} = 2\times10^{21} \rm{cm}^{-2}$, the relative abundance of DCO$^+$ to HCO$^+$ is as high as 0.3 within the sub-region which lies near 20 K.  This represents a local enhancement by a factor of at least 10$^4$ over the ISM [D]/[H] ratio of $\approx$10$^{-5}$ \citep{Watson:1976}, and is a higher deuteration than that observed in pre-stellar cores where CO is also frozen out \citep[e.g.,][]{Caselli:1999}.  Some models of vertical chemical structure predict that the HCO$^+$ abundance begins declining above the region of enhanced DCO$^+$ abundance \citep{Aikawa:2006}.  In such a scenario, the local DCO$^+$ to HCO$^+$ abundance ratio would be even higher.  While H$^{13}$CO observations could provide constraints on this scenario, distentangling the midplane emission from that in higher layers in the disk requires more detailed modelling of the HCO$^+$ abundance structure and higher signal-to-noise H$^{13}$CO$^+$ data.

In the images presented here, DCO$^+$ emission covers approximately 1/4 of the observed HCO$^+$ emission area (Fig. \ref{fig:allLines}).  In our model, the vertical thickness of the DCO$^+$ region is also approximately 1/4 that of the HCO$^+$ region (Fig. \ref{fig:abundance}).  Taking into account both geometric effects, the disk averaged abundance ratio is approximately 0.02, comparable to the disk-averaged abundance ratio found in TW Hya \citep[0.04,][]{2003A&A...400L...1V}.

\subsection{Ionization fraction}
The model HCO$^+$ and DCO$^+$ abundances provide a lower limit to the ionization fraction of the disk.  In the warm molecular disk where HCO$^+$ is present, the ionization fraction must be at least $3\times10^{-10}$.  Assuming that it continues to be present into the cold midplane region where the DCO$^+$ abundance is enhanced (as it is in our simple model), the minimum ionization fraction would be $4\times10^{-10}$ (HCO$^+$ + DCO$^+$).  This is at least an order of magnitude higher than that found in the cool molecular layer of DM Tau described by \cite{Oberg:2011c}.  This is also three orders of magnitude higher than the critical ionization fraction at which dead zones form.  Observations of other region specific ions \citep[e.g., H$_2$D$^+$ outside the CO snowline,][]{Oberg:2011c} will allow for estimation of the importance of this mechanism for planet formation throughout the disk. 

\subsection{Comparison to other disks}

The scenario we have modeled here for HD 163296 is consistent with that described for DM Tau by \cite{Oberg:2011c}, wherein they find that unresolved DCO$^+$ emission can be fit using an enhanced DCO$^+$ abundance in the temperature range of 16-20 K.  This temperature range is slightly lower and broader than that found here.

DCO$^+$ $J$=3-2 emission has been mapped towards TW Hya by \cite{Qi:2008} and can also be fit by a ring-like morphology.   They simulate a three-layer disk structure by assuming a model in which the molecular column density is a function of distance from the star, and the abundance at any given radius is constant within the middle layer.  They fit a power law radial column density profile for each molecule in their study, as well as a vertical structure defined relative to the hydrogen column density.  Under this model, they find that DCO$^+$ emission is best fit by either a model in which its column density increases away from the star with a sharp truncation at 90 AU, or by a two-part power law function which increases to 70 AU, and then decreases.  In both models, the DCO$^+$ exists in a layer extending from hydrogen column densities of 0.1 to 10 times $1.59\times10^{21}$ cm$^{-2}$, the `disk surface'.  This region has temperatures ranging from $\sim$16--50 K, and densities from $\sim10^7$--$10^8$ cm$^{-3}$.  Midplane temperatures at this radius are $\sim10$ K.

The 19 K midplane isotherm in TW Hya lies at 30 AU in the \cite{Qi:2008} model. DCO$^+$ in this disk therefore does not appear to trace the same isotherm as in HD 163296 and in pre-stellar cores.  Either some mechanism acts to inhibit the low temperature enhancement of DCO$^+$ at 30 AU, or alternatively, some process generates a second peak in DCO$^+$ around 70 AU as seen by the SMA.

One possibility is related to the formation of H$_2$D$^+$, the DCO$^+$ parent molecule.  Whereas we have focused on the reduction of CO abundance allowing for greater formation of H$_2$D$^+$ from H$_3^+$, other molecules can serve to inhibit its formation (therefore inhibiting its reaction products, such as DCO$^+$).  H$_3^+$ can also react with N$_2$.
Perhaps the 70 AU radius in TW Hya traces N$_2$ freezeout which occurs at slightly lower temperatures \citep{Bisschop:2006fk} resulting in a second peak in H$_2$D$^+$.  If some residual CO was present, due to mixing from warmer layers or release from grains, then DCO$^+$ could form.

Alternatively, the underlying model of gas and dust density and temperature could be in need of revision.  Recent SMA continuum observations \citep{Andrews:2012} indicate the mm-emitting dust disk is truncated at 60 AU.  This would have far reaching effects on the thermal structure and disk chemistry, including the potential release of CO into the gas phase due to increased penetration of ultraviolet radiation (an effect noted by Andrews et al.).  This could allow the gas-phase formation of DCO$^+$ even outside the CO snowline.

At 30 AU, it is also possible that the ortho-to-para ratio of H$_2$ has not dropped to the low levels of $\sim10^{-3}$ expected for these low midplane temperatures and required to enhance H$_2$D$^+$ abundance. The ortho-to-para conversion is governed by the H$^+$ + ortho-H$_2$ $\to$ H$^+$ + para-H$_2$ reaction, which has a timescale that is proportional to the H$^+$ abundance.  For an H$^+$ abundance of at most $10^{-12}$ with respect to H$_2$ \citep{Walsh:2012qy} the timescale is at least $3\times10^5$ yr for a density of $10^9$ cm$^{-3}$ and rate coefficient of $1.1\times 10^{-10}$ cm$^3$ s$^{-1}$ at 10 K \citep{Pagani:2013uq}.  If some high temperature event (e.g., a recent flare) has reset the H$_2$ ortho/para ratio to a high temperature value in the inner disk temporarily, it may not have equilibrated back to the low value needed to boost the H$_2$D$^+$ abundance at the 30 AU radius. Note that the ortho-H$_2$ S(1) line at 17 $\mu$m has been detected toward TW Hya \citep{Najita:2010lr} and interpreted to arise from radii out to 30 AU \citep{Gorti:2011fj}.  Even though the mid-infrared emission arises from the warm surface layers, some of this ortho-H$_2$ may have been mixed down to the midplane.

A superthermal abundance of ortho-H$_2$ also has other chemical consequences, most notably in driving the reaction N$^+$ + ortho-H$_2$ $\to$ NH$^+$ + H, which triggers the nitrogen chemistry leading to NH$_3$.  The strong {\it Herschel}-HIFI detection of NH$_3$ in the TW Hya disk (Hogerheijde et al., in prep) would be consistent with this scenario.

Ultimately, this discussion points to the need for high spatial resolution ALMA mapping of DCO$^+$ and other molecules sensitive to CO freezeout \citep[e.g.,][]{Qi:2013fk} in the TW Hya and other disks to determine the relative importance of the H$_2$D$^+$ enhancement versus CO freeze-out.

\subsection{Dependence on physical models}

Here we relate the observed emission to physical/chemical properties via an adopted structure model for the disk. This model has been constrained via observations that trace various regions of the disk (e.g., the SED and CO observations).  CO observations in particular are primarily sensitive to a thin surface layer between the height where photodissociation of CO no longer effectively operates and the height where the lines become optically thick. Even the emission of optically thin isotopologues of CO are strongly affected by the emission from these surface layers, and only partially probe the disk midplane.  Because our DCO$^+$ observations are thought to probe the disk midplane, it is appropriate to investigate the dependence of our results on the details of the adopted physical model.

Compared to our model, which is based on the work of Q11, the best-fit disk structure of de Gregorio-Monsalvo et al. (in press) has 1--3 K lower temperatures and factor $\sim1.5$ higher densities in the 100 -- 150 AU midplane region. Using this model, we find that our observations are best described if DCO$^+$ is enhanced between temperatures of 16 and 20 K (compared to 19--21 K found using the Q11 model). The best fit DCO$^+$ abundance is $\sim7\times10^{-11}$, compared to $1\times10^{-10}$ for the Q11 model. These higher densities affect the H$^{13}$CO$^+$ abundance in a similar fashion, leading to an HCO$^+$ abundance of $\sim 2\times10^{-10}$.  The differences in gas density affect the molecular abundance in the same manner, leading to a similar localized DCO$^+$ / HCO$^+$ ratio of $\sim0.3$.

Although the details of the best-fit model clearly depend on the adopted disk structure, the general characteristics remain unchanged: DCO$^+$ is a sensitive tracer of the narrow temperature range where CO freezes out and H$_2$D$^+$ starts to increase in abundance. Other high-resolution observations of species that also trace this transition (e.g., N$_2$H$^+$; see Qi et al., in prep.) will further constrain the parameter range of allowed models.

\section{Conclusions}
\label{sec:conclusions}

We have presented ALMA Science Verification images of the HCO$^+$ $J$=4$-$3, H$^{13}$CO$^+$ $J$=4$-$3, and DCO$^+$ $J$=5$-$4 emission from the \mbox{HD 163296} circumstellar disk.  While HCO$^+$ traces the entire disk and is detected at the stellar position, the DCO$^+$ emission is seen in a ring tracing the region between the rise of H$_2$D$^+$ formation and the CO snowline.  Using radiative transfer modelling, we have shown that the emission can be matched by a model in which DCO$^+$ has high abundance within a limited temperature range near the CO freezeout temperature (from 19 to 21 K).  This emission traces the previously determined CO snowline.  Comparison of the modeled abundances indicates a [DCO$^+$] / [HCO$^+$] ratio of 0.3, the highest yet observed.  This in turn suggests that gas phase deuteration processes can lead to local enhancements of the abundance of DCO$^+$ on the order of a factor of 10$^4$, comparable to that observed in some dark clouds.  The high abundance of these ions also indicates that the ionization fraction must be $\gtrsim10^{-10}$ in the cold midplane, three orders of magnitude higher than necessary to maintain the MHD turbulance that powers viscous accretion.

Future high sensitivity, high resolution observations with ALMA will allow for the direct imaging of many chemical processes in disks.  
These studies will provide important constraints on models of disk physical structure even in the vertical direction.  In conjunction with tracers of the disk surface (e.g. CO) and the bulk gas (e.g. C$^{18}$O), we are entering the era of `disk tomography' where disks can be mapped in a region-by-region fashion.  This will, in turn, allow us to study  chemistry and dynamics as they vary throughout the disk.

\begin{acknowledgements}
  %  ALMA Acknowledgement information from http://www.nrao.edu/library/pagecharges.shtml 
  This paper makes use of the following ALMA data: ADS/JAO.ALMA\#2011.0.00010.SV. ALMA is a partnership of ESO (representing its member states), NSF (USA) and NINS (Japan), together with NRC (Canada) and NSC and ASIAA (Taiwan), in cooperation with the Republic of Chile.  The Joint ALMA Observatory is operated by ESO, AUI/NRAO and NAOJ.  
  Astrochemistry in Leiden is supported by NOVA, KNAW and EU A-ERC grant 291141 CHEMPLAN.  
  D.H. is additionally supported by SRON.
  IdGM acknowledges the Spanish MICINN grant AYA2011-30228-C03 (co-funded with FEDER funds).
  AH acknowledges support from the Millennium Science Initiative, Chilean Ministry of Economy: Nucleus P10-022-F.
  The National Radio Astronomy Observatory is a facility of the National Science Foundation operated under cooperative agreement by Associated Universities, Inc.

\end{acknowledgements}

\bibliographystyle{aa}
%\bibliography{HD163296_DCOp_bib.bbl}

\clearpage
\newpage

\Online
\appendix

\section{Modeling details}
\label{sec:model_details}

\subsection{Parametric dust and gas structure}
We adopt the overall disk properties derived by Q11 from fitting the broadband spectral energy distribution and spatial extent of their mm data: a disk mass $M_{\rm{disk}}$ = 0.089 \Msun, with an inner edge $R_{\rm{in}}$ = 0.6 AU and a critical radius $R_{\rm{C}}$ = 150 AU.  We define the surface density as \citep[e.g.][]{2009ApJ...700.1502A}:
\begin{equation}
  \Sigma (R) = \Sigma_C \left ( \frac{R}{R_{\rm{C}}} \right ) ^{-\gamma} \exp \left [-\left (\frac{R}{R_{\rm{C}}} \right )^{2-\gamma} \right ], 
\end{equation}
where we set the surface density power law index, $\gamma$, equal to 1, the value for a self-similar accretion disk in steady-state \citep[][]{Hartmann:1998}.  We adopt the gas-to-dust ratio of 154 from Q11 as well as their two component dust grain population (small and large) used to simulate dust settling.  Both populations are made of a 60\%/40\% mix of astronomical silicates and graphite, and follow a power law grain size distribution $n(a) \propto a^{-3.5}$ \citep{Draine:2006} with a minimum grain size $a_{min} = 0.005$ $\mu$m for both populations.  The small grain population has a maximum grain size $a_{\rm{max, small}}=0.25$ $\mu$m, and the large population has $a_{\rm{max, large}}=1$mm.  The small grain component is set to include 1.6\% of the dust mass ($f_{\rm{small}}$ = 0.016). 

The vertical distribution is modeled as a Gaussian with angular scale height
\begin{equation}
h=h_C \left (\frac{R}{R_{\rm{C}}} \right )^{\psi}, 
\end{equation}
where $h_{\rm{C}}$ is the angular scale height at the characteristic radius $R_{\rm{C}}$ and $\psi$ describes the power-law disk flaring.  We set $\psi$ = 0.066, for consistency with the modeling of \cite{Tilling:2012}.  The two dust component populations have independent scale heights ($h_{\rm{C}}$) of $h_{\rm{small}}$ and $h_{\rm{large}}$, respectively, which we adjust to match the spectral energy distribution (discussed below). 

The gas density distribution produced by the physical model of Q11 is approximately gaussian at low heights and has a long `tail' to large heights (c.f., their Figure 7).  In order to approximate this, we have constructed the gas density distribution using a two-component model with independently varying scale heights for the main component ($h_{\rm{main}}$) and the lower mass, larger scale height tail ($h_{\rm{tail}}$).  In addition, we vary the distribution of gas mass between these two components, with a term $f_{\rm{tail}}$ describing the fraction of the total gas mass in the tail.   

\subsection{Determining the disk parameters}

\begin{figure}
   \centering
   \includegraphics[angle=0, width=\columnwidth]{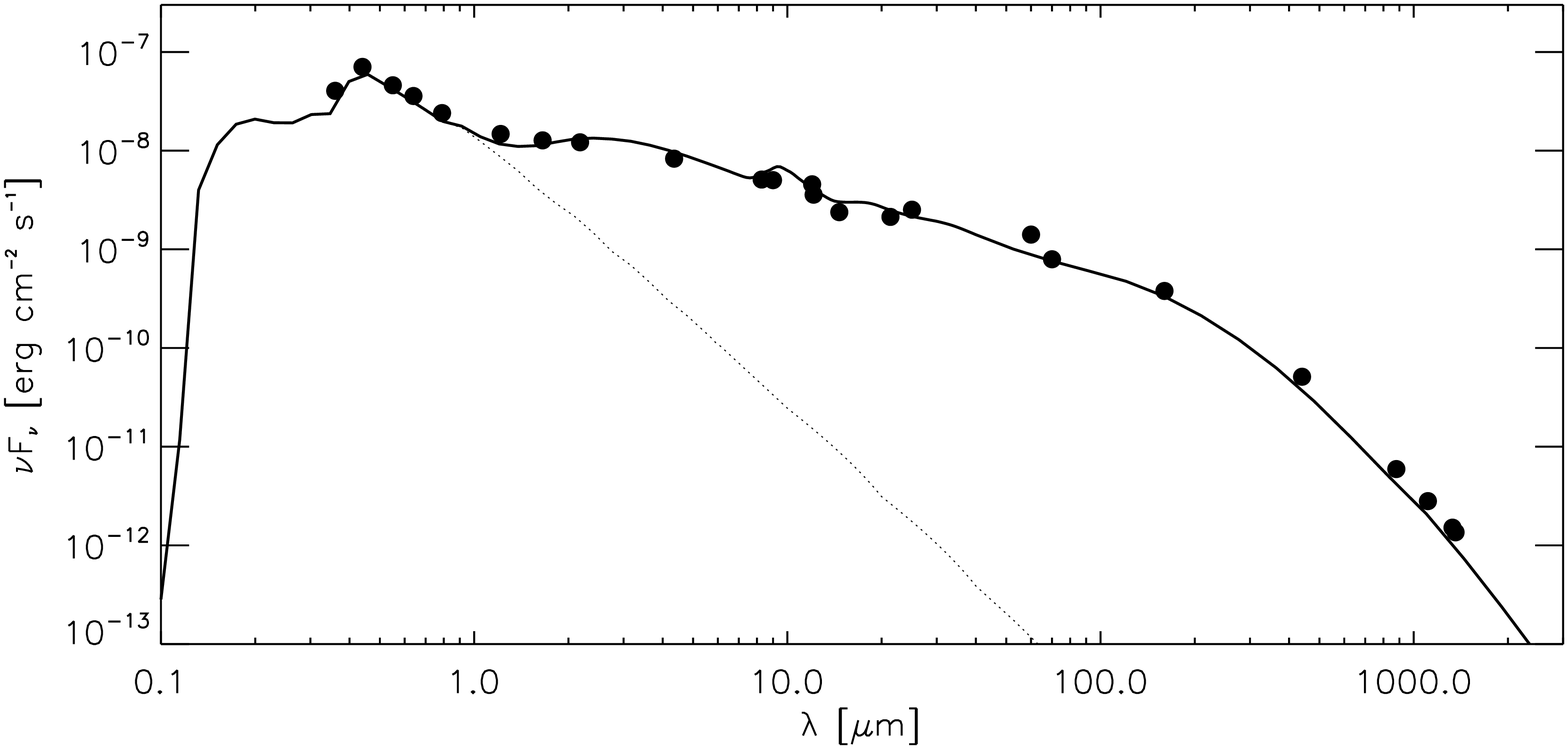}
      \caption{Spectral energy distribution (black line) of our best fit model, compared to dereddened photometry ($A_V$=0.3, Q11) from the literature \citep[Q11,][]{Tilling:2012, Cutri:2011, Ishihara:2010, Egan:2003,1988iras....7.....H}.  The stellar photosphere is shown as the thin dashed line.}
         \label{fig:SED}
   \end{figure}

With this framework in place, we adjust parameters to match the SED and CO line fluxes in the literature for \mbox{HD 163296}.  
We carried out continuum radiative transfer modeling using the 2D code RADMC \citep{2004A&A...417..159D}, which receives as input stellar properties and a dust density structure and outputs the resulting temperature structure.  It includes a ray-tracing code, Raytran, for producing the model spectral energy distribution. 
We adjusted $h_{\rm{small}}$ ~and $h_{\rm{large}}$ in order to approximate the temperature structure found in Q11 (c.f. their Figure 9), which came from a self-consistent vertical structure calculation.  We made further small adjustments to these parameters in order to match the SED.  Our adopted disk model has small- and large-grain scale heights of 0.08 and 0.06, respectively, and we show the resulting SED in Figure \ref{fig:SED}.  In our further modeling, we assume the gas temperature matches the mean dust temperature calculated in this modeling, which is a reasonable approximation in the dense regions upon which we focus.  

After setting the dust parameters, we adjust $h_{\rm{main}}$, $h_{\rm{tail}}$, and $f_{\rm{tail}}$ to approximate the gas density structure found in Q11 (their Figure 9).  
We fine-tune these parameters by comparing our modeled fluxes of CO and its isotopologues with the observed fluxes of Q11, focusing on the low-$J$ rotational lines that are expected to come largely from the cool outer disk regions studied here.  
We assume a relative CO abundance [CO] / [H] $\approx$ 10$^{-4}$ \citep{Frerking:1982}, which drops to zero at hydrogen column densities less than $2\times10^{21}$ cm$^{-2}$ due to photodissociation \citep[][]{Visser:2009a} and at temperatures less than 19 K due to freezeout \citep{Qi:2011}.  We also assume a ratio of $^{12}$C to $^{13}$C of 75, and ratios of $^{16}$O to $^{18}$O and $^{17}$O of 500 and 1750, respectively \citep{Frerking:1982}. 
To generate our model line emission, we input our adopted dust density and temperature structure, along with the model gas density structure, to the line radiative transfer code LIME \citep{Brinch:2010}, and we achieve a match to the observed line fluxes within $\sim$50\%.

 The gas `tail' has a scale height $h_{\rm{tail}}$ = 0.2 and includes 5\% of the gas mass ($f_{\rm{tail}}$), while the `main' gas component has a scale height $h_{\rm{main}}$ = 0.1.   
In Table \ref{tab:modelfluxes}, we list the integrated fluxes of the CO lines which we use to determine the disk gas structure.  
Table \ref{tab:model} lists and categorizes all disk modeling parameters, as well as the range of explored values or the reference from which the fixed value was adopted.  In both tables, we repeat the HCO$^+$, H$^{13}$CO$^+$, and DCO$^+$ properties discussed in the main body of the text.

\begin{table}[th]
     \center
      \caption[]{Comparison of observed and model fluxes}
      \label{tab:modelfluxes}
%       \resizebox{18.5cm}{!} { 

         \begin{tabular}{l c c  }
            \hline\hline
            \noalign{\smallskip}
            Property   										&  Model  							&  Observed  \\
            		   										&  (Jy km s$^{-1}$)  					&   (Jy km s$^{-1}$) \\
            \noalign{\smallskip}
            \hline
            \noalign{\smallskip}	%											mod00471
		CO \hspace{2.05em}$J$=2$-$1					&  	51\phantom{.0}					& 	 54.17$\pm$0.39\tablefootmark{a}	\\%	
		$^{13}$CO \hspace{1.25em}$J$=2$-$1				&  	21\phantom{.0}					& 	 18.76$\pm$0.24\tablefootmark{a}	\\%		
		C$^{18}$O \hspace{1.25em}$J$=2$-$1				&  	\phantom{0}9\phantom{.0}			& 	 \phantom{0}6.30$\pm$0.16\tablefootmark{a}	\\%		
		CO \hspace{2.1em}$J$=3$-$2						&  	74\phantom{.0}					& 	 98.72$\pm$1.69\tablefootmark{a}\\%		
		C$^{17}$O \hspace{1.25em}$J$=3$-$2				&  	\phantom{0}7\phantom{.0}			& 	 11.64$\pm$0.76\tablefootmark{a}	\\%	
		HCO$^+$ \hspace{0.75em}$J$=4$-$3				&  	26.0							& 	18.7$\pm$0.7\phantom{0}	 \\%			
		H$^{13}$CO$^+$ $J$=4$-$3						&  	\phantom{0}1.4					& 	\phantom{0}1.0$\pm$0.2\phantom{0}	  \\%			
		DCO$^+$ \hspace{0.7em}$J$=5$-$4				&  	\phantom{0}1.7					& 	\phantom{0}2.2$\pm$0.4\phantom{0}	 \\%			
           \noalign{\smallskip}
            \hline
         \end{tabular}
           \\
           
           \tablefoot{
             \tablefoottext{a} {Observed} fluxes from Q11.
           }
 \end{table}

\begin{table}[th]
     \center
      \caption[]{Adopted model}
      \label{tab:model}
%       \resizebox{18.5cm}{!} { 

         \begin{tabular}{lll  }
            \hline\hline
            \noalign{\smallskip}
            Property   					&  Value  			&  Explored range \\
            							&				&  or reference  \\
            \noalign{\smallskip}
            \hline
            \noalign{\smallskip}	
            	Stellar parameters			&				&		\\
	  \hline
		$M_{\rm{star}}$				&  2.3 \Msun		&  Q11	\\
		$R_{\rm{star}}$				&  2.0 \Rsun		&  Q11	\\
		$T_{\rm{star}}$				&  9333 K			&  Q11	\\
	\hline
		Dust parameters			&				&		\\
	\hline
		[silicate] / [graphite]			&  60\% / 40\%		&  Q11	\\
		$a_{\rm{min}}$				&  0.005 $\mu$m	&  Q11	\\
		$a_{\rm{max, small}}$		&  0.25 $\mu$m	&  Q11	\\
		$a_{\rm{max, large}}$		&  1 mm			&  Q11	\\
		$f_{\rm{small}}$			&  0.016			&  Q11	\\
	\hline
		Disk parameters			&				&		\\
	\hline
		$M_{\rm{disk}}$			&  0.089 \Msun		&  Q11	\\
		gas/dust					&  154			&  Q11	\\	
		$R_{\rm{in}}$				&  0.6 AU			&  Q11	\\
		$R_C$					&  150 AU			&  Q11	\\
		$h_{\rm{puffed rim}}$		&  1.67			&  Q11	\\
		$\gamma$				&  1				&  \cite{Hartmann:1998}	\\
		$\psi$					&  0.066			&  \cite{Tilling:2012}	\\
		Inclination					&  44\degr			&  Q11	\\
		Position angle				& 133\degr 		&  Q11	\\
		Systemic velocity			&  5.8 km s$^{-1}$		& Q11	\\
		$f$(\rm{Small grains})		&  0.016			&  Q11	\\
		$h_{\rm{small}}$			&  0.08			&  0.04 -- 0.20 \\
		$h_{\rm{large}}$			&  0.06			&  0.04 -- 0.20 \\
		$h_{\rm{main}}$			&  0.10			&  0.04 -- 0.24 \\
		$h_{\rm{tail}}$				&  0.20			&  0.04 -- 0.24 \\
		$f_{\rm{tail}}$				&  0.05			&  0.00 -- 0.20 \\
	\hline
		Molecule parameters		&				&		\\
	\hline
		\raiseChi$_{\rm{CO}}$					&  10$^{-4}$		&  \cite{Frerking:1982}	\\
 		\raiseChi$_{\rm{HCO^+}}$				&  $3\times10^{-10}$ & 10$^{-11}$ -- 10$^{-5}$ \\
		\raiseChi$_{\rm{H^{13}CO^+}}$			&  \raiseChi$_{HCO^+}$ / 75	& \cite{Frerking:1982} \\
		\raiseChi$_{\rm{DCO^+}}$				&  $1\times10^{-10}$ & 10$^{-11}$ -- 10$^{-8}$ \\
		$T_{\rm{high, DCO^+}}$			&  21K			& 19 -- 25K \\
		$T_{\rm{low, DCO^+}}$			&  19K			& 16 -- 21K \\

           \noalign{\smallskip}
            \hline
         \end{tabular}
           \\
        
 \end{table}

\end{document}